\documentclass[prd,twocolumn,showpacs,floatfix,amsmath,amssymb,floatfix,nofootinbib]{revtex4}
\usepackage{graphicx,color,dcolumn,booktabs,bm}
\usepackage{amsmath}
\usepackage{longtable,lscape,comment,braket}
\usepackage{txfonts}
\usepackage{overpic}
\usepackage{indentfirst}
\usepackage{feynmf}   
\usepackage{slashed}  
\usepackage{cases}
\usepackage{epstopdf}
\usepackage{psfrag}
\usepackage{subfigure}
\usepackage{threeparttable}
\pdfoutput=1
\usepackage{color}
\usepackage{multirow}

\def\lrpartial{\buildrel\leftrightarrow\over\partial}
\graphicspath{{Figures/}} %
\usepackage[colorlinks, citecolor=blue,anchorcolor=red,menucolor=red, linkcolor=red,filecolor=red,runcolor=red,urlcolor=blue,frenchlinks=red]{hyperref}

\usepackage{ulem}
\def\be{\begin{equation}}
\def\ee{\end{equation}}

\begin{document}

\title{Proposal of searching for the $\Upsilon(6S)$ hadronic decays into $\Upsilon(nS)$ plus $\eta^{(\prime)}$}
\author{Qi Huang$^{1,2}$}
\author{Xiang Liu$^{1,2}$}\email{xiangliu@lzu.edu.cn}
\author{Takayuki Matsuki$^{3,4}$}\email{matsuki@tokyo-kasei.ac.jp}
\affiliation{$^1$School of Physical Science and Technology, Lanzhou University, Lanzhou 730000, China\\
$^2$Research Center for Hadron and CSR Physics, Lanzhou University and Institute of Modern Physics of CAS, Lanzhou 730000, China\\
$^3$Tokyo Kasei University, 1-18-1 Kaga, Itabashi, Tokyo 173-8602, Japan\\
$^4$Theoretical Research Division, Nishina Center, RIKEN, Wako, Saitama 351-0198, Japan}

\begin{abstract}

In this work, we propose a possible experimental research topic on the $\Upsilon(6S)\to \Upsilon(nS)\eta^{(\prime)}$ ($n=1,2,3$) transitions.
Considering the hadronic loop effect from the intermediate $S$-wave $B_{(s)}$ and $B_{(s)}^*$ mesons, we estimate the partial decay widths and the corresponding branching ratios of the $\Upsilon(6S)\to \Upsilon(nS)\eta^{(\prime)}$ hadronic decays, which become considerably large. With the running of the Belle II experiment, it becomes possible to explore these $\Upsilon(6S)\to \Upsilon(nS)\eta^{(\prime)}$ transitions.
\end{abstract}


\maketitle

\section{introduction}\label{sec1}

Since 2008, experimentalists have made a great progress in exploring the hadronic decays of higher bottomonia.  There exist several  typical examples: (1) The measured partial widths of the $\Upsilon(10860) \to \Upsilon(mS) \pi^+ \pi^-~(m\leq3)$ transitions are around $\mathcal{O}(1)$ MeV, which are anomalous since the $\Upsilon(10860) \to \Upsilon(1S) \pi^+ \pi^-$ is at least 2 orders of magnitude larger than those of $\Upsilon(nS) \to \Upsilon(1S) \pi^+ \pi^-~(n\leqslant4)$ \cite{Abe:2007tk}. (2) The experimental branching ratios of the $\Upsilon(5S) \to \chi_{bJ} \omega~(J=0,1,2)$ transitions can reach up to $10^{-3}$ \cite{He:2014sqj}. (3) Two charged bottomoniumlike structures $Z_b(10610)$ and $Z_b(10650)$ were discovered by Belle when analyzing hidden-bottom dipion decays of the $\Upsilon(10860)$ \cite{Belle:2011aa,Krokovny:2013}.

For solving these puzzling phenomena relevant to the hadronic decays of the $\Upsilon(10860)$, the hadronic loop mechanism
was introduced \cite{Chen:2011zv,Meng:2007tk,Meng:2008dd,Meng:2008bq,Simonov:2008qy,Chen:2011qx,Chen:2014ccr,Chen:2011pv,Wang:2016qmz}, and equivalent to the description of the coupled channel effect. Recently, the Belle Collaboration announced a new hidden-bottom decay mode $\Upsilon(5S) \to \Upsilon(1^3D_J)\eta$, the branching ratios of which are around $10^{-3}$ \cite{Tamponi:2018cuf}. This observation directly confirmed the former theoretical prediction given in Ref. \cite{Wang:2016qmz}. Very recently, Belle also observed $\Upsilon(6S)\to \chi_{bJ}\pi^+\pi^-\pi^0$ and measured its branching ratio $B(\Upsilon(6S)\to \chi_{bJ}\pi^+\pi^-\pi^0)=(8.6\pm4.1\pm6.1^{+4.5}_{-2.5})\times 10^{-3}$ \cite{Yin:2018ojs}, which is consistent with the theoretical prediction by Ref. \cite{Huang:2017kkg}. These facts further show the success of the hadronic loop mechanism involved in the hadronic decays of higher bottomonia.

\begin{center}
	\begin{figure*}[htbp]
		\scalebox{0.5}{\includegraphics{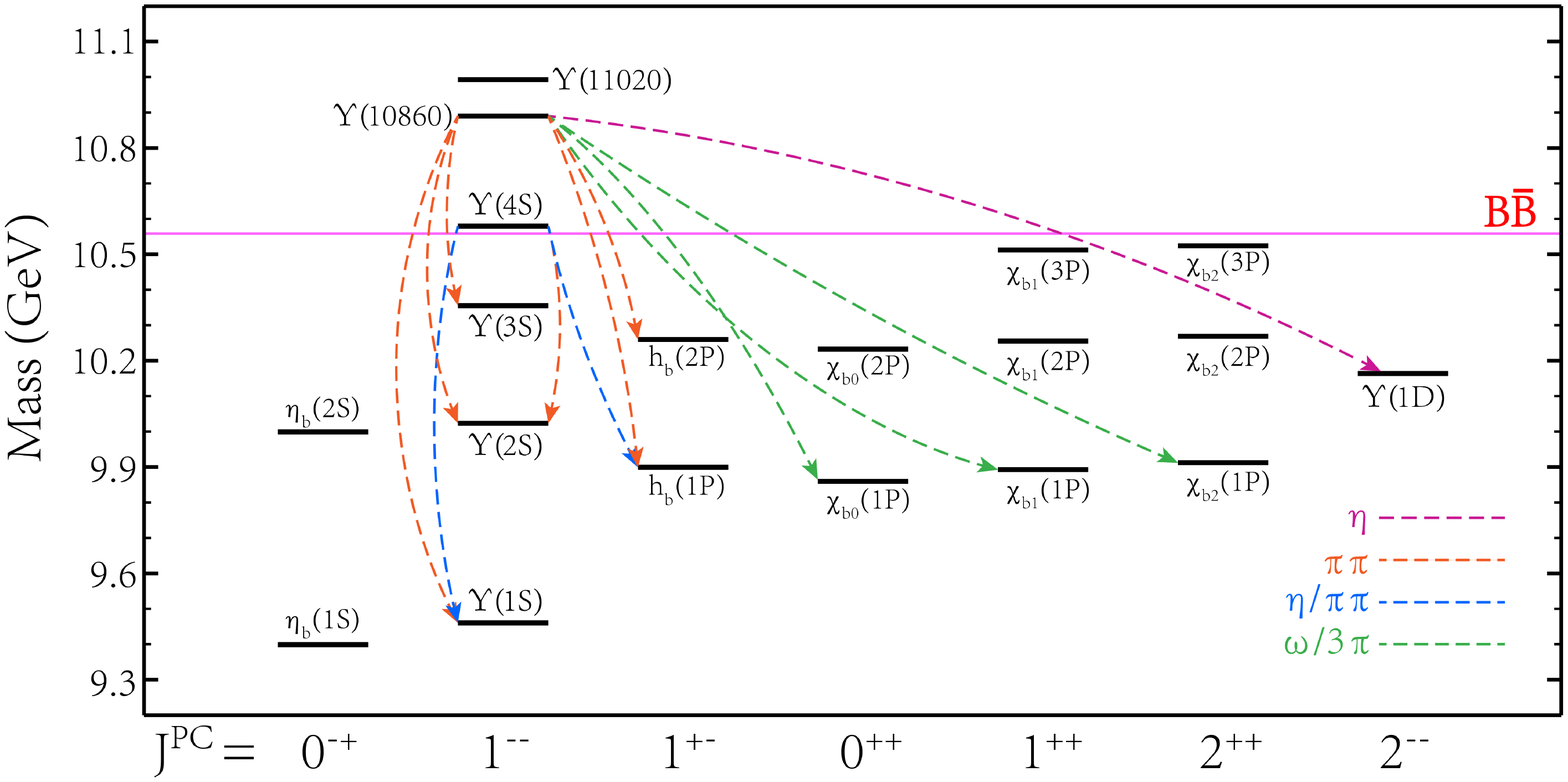}}
		\caption{The observed bottomonia and their allowed hadronic transitions \cite{Olive:2016xmw}.}
		\label{fig:spectrum}
	\end{figure*}
\end{center}

We notice that the Belle II experiment measured the first $e^+e^-$ collision on April 26, 2018, which was an exciting moment.
With the running of Belle II, we have reason to believe that it will become an interesting research topic at Belle II to explore new hadronic decay modes of higher bottomonia.
Just shown in Fig. \ref{fig:spectrum}, there are six observed bottomonia, i.e., $\Upsilon(1S)$, $\Upsilon(2S)$, $\Upsilon(3S)$, $\Upsilon(4S)$, $\Upsilon(10860)$, and $\Upsilon(11020)$, where the last two states are usually treated as 5S and 6S states of the $\Upsilon$ family\footnote{In the following, $\Upsilon(11020)$ is abbreviated to $\Upsilon(6S)$.}. Since the mass of $\Upsilon(11020)~(\Upsilon(6S))$ is higher than that of the $\Upsilon(10860)~(\Upsilon(5S))$, we may naturally infer that the hadronic loop mechanism may still play an important role in the hadronic decays of the $\Upsilon(6S)$ according to the experience of studying the $\Upsilon(5S)$ transitions and $\Upsilon(6S)$ \cite{Wang:2016qmz,Huang:2017kkg,Huang:2018cco}. In Refs. \cite{Huang:2017kkg,Huang:2018cco}, we estimated the branching rates of the $\Upsilon(6S)$ decays into $\chi_{bJ}\omega$, $\chi_{bJ}\phi$ \cite{Huang:2017kkg}, and $\Upsilon(1^3D_J)\eta$ \cite{Huang:2018cco}.

Along this line, in this work we continue to focus on the
specific hadronic transitions $\Upsilon(6S)\to \Upsilon(nS)\eta^{(\prime)}$ ($n=1,2,3$).  Using the hadronic loop mechanism, we need to further estimate their partial decay widths and the corresponding branching ratios, which are crucial information for experimental analysis of these decays at Belle II.
In addition, we predict several typical ratios of these discussed hadronic decays, which can be tested in future experiment.

Frankly speaking, another motivation inspiring us to carry out such a study is that the present work is one part of the whole physics around the higher bottomonium $\Upsilon(6S)$. We hope that the entire aspect of the $\Upsilon(6S)$ can be described by our step-by-step effort, which will be valuable to in determining these potential research topics at Belle II when more and more data will be accumulated in the near future.

This work is organized as follows. After the Introduction, we spend the next section illustrating the detailed calculation of the $\Upsilon(6S)\to \Upsilon(nS)\eta^{(\prime)}$ transitions (see Sec. \ref{sec2}). Then, the numerical results including the partial decay widths, the branching ratios, and several ratios will be presented in Sec. \ref{sec3}. This paper ends with conclusions and a discussion.

\section{Estimate of the decay rates $\Upsilon(6S)\to \Upsilon(nS)\eta^{(\prime)}$}\label{sec2}

\begin{center}
\begin{figure}[htbp]
\scalebox{0.3}{\includegraphics{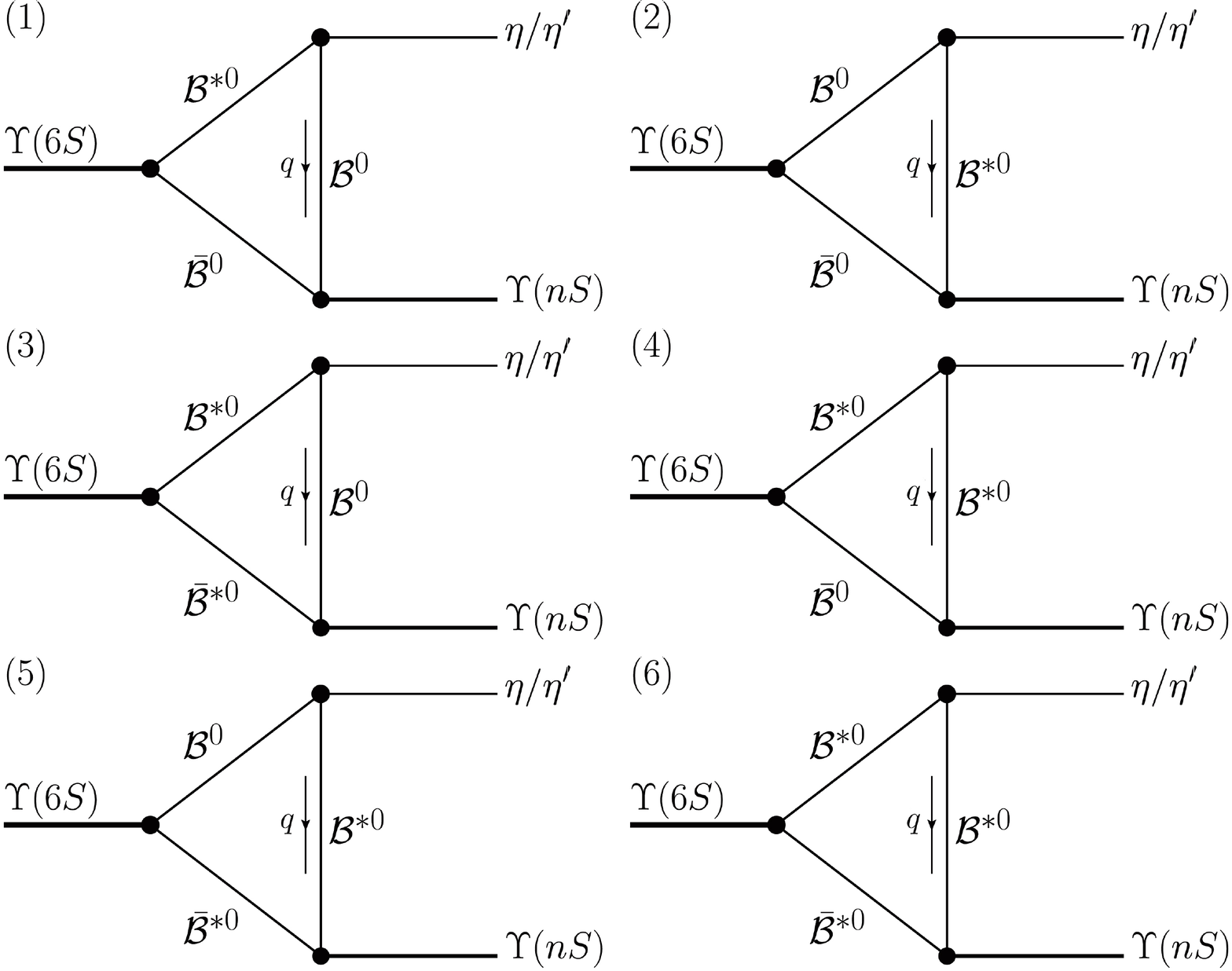}}
\caption{The involved diagrams of the $\Upsilon(6S) \to \Upsilon(nS) \eta^{(\prime)}~(n=1,2,3)$ transitions via the hadronic loop mechanism.}
\label{fig:6S-nS-eta-all}
\end{figure}
\end{center}

Since the $\Upsilon(6S)$ dominantly decays into a pair of bottom or bottom-strange mesons, under the framework of the hadronic loop mechanism, the hidden-bottom decays of the $\Upsilon(6S)$ can be achieved in the following way. First, the $\Upsilon(6S)$ is decomposed into $B_{(s)}^{(*)}\bar{B}_{(s)}^{(*)}$, and then by exchanging a $B_{(s)}^{(*)}$ meson, the bottom mesons convert themselves into the $\Upsilon(nS)$ and $\eta/\eta'$. The Feynman diagrams that depict this process are shown in Fig. \ref{fig:6S-nS-eta-all}.

\begin{center}
	\begin{figure}[htbp]
		\scalebox{0.4}{\includegraphics{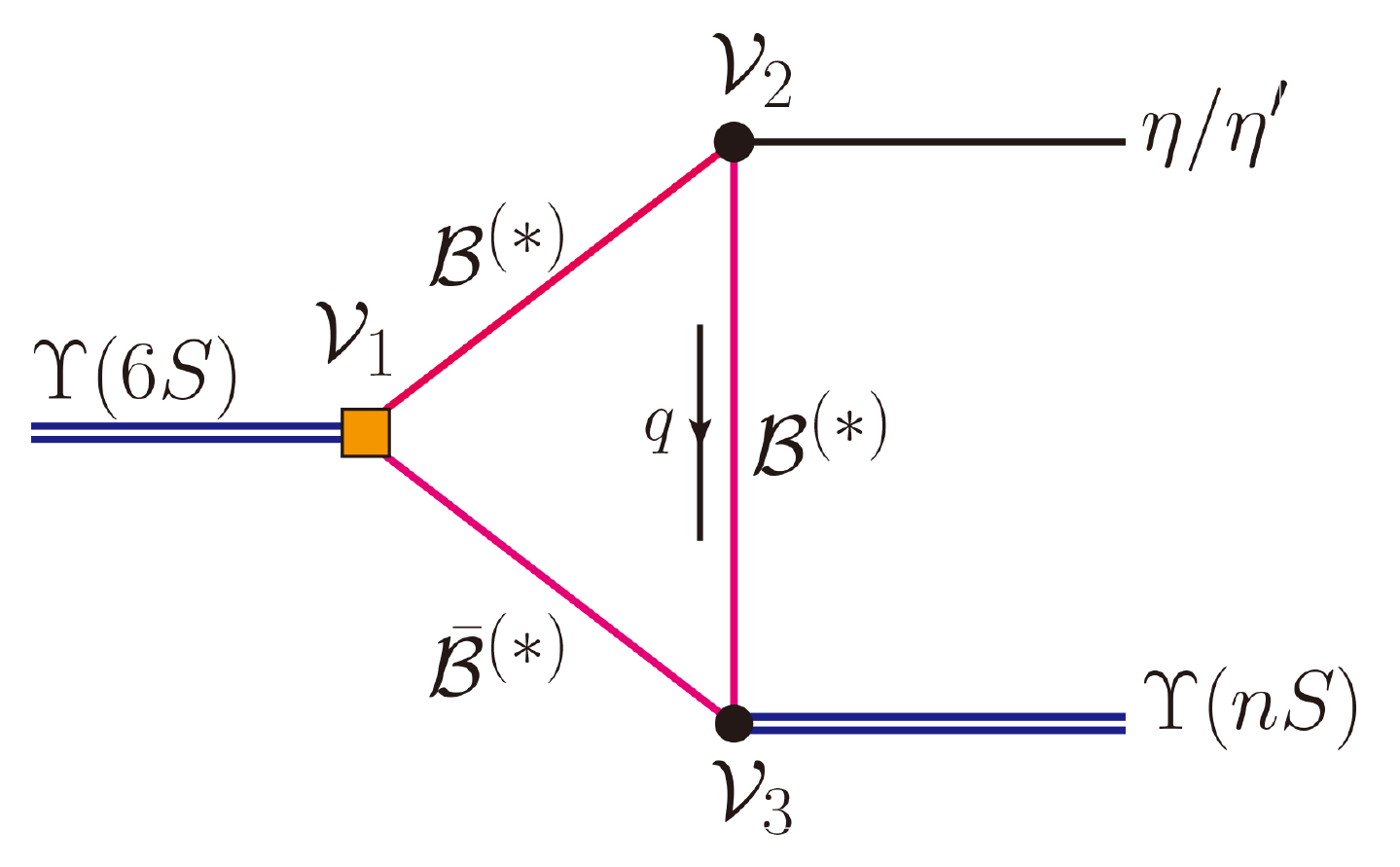}}
		\caption{The schematic diagram depicting the transitions $\Upsilon(6S)\to\Upsilon(nS)+\eta/\eta'$.}
		\label{fig:6S-nS-eta}
	\end{figure}
\end{center}

All the diagrams shown in Fig. \ref{fig:6S-nS-eta-all} can be abstracted into one general diagram presented in Fig. \ref{fig:6S-nS-eta}, 
in which the bottom or bottom-strange mesons construct the hadronic loop and connect the initial $\Upsilon(6S)$ with the final $\Upsilon(nS)$ and $\eta^{(\prime)}$ so that the general expression of the amplitude, according to Fig. \ref{fig:6S-nS-eta}, then can be written as
\begin{eqnarray}
\mathcal{M}=\int \frac{d^4q}{(2\pi)^4}\frac{\mathcal{V}_1\mathcal{V}_2\mathcal{V}_3}{\mathcal{P}_1 \mathcal{P}_2 \mathcal{P}_{E}}\mathcal{F}^2(q^2, m_E),\label{eqs:6S-nS-eta}
\end{eqnarray}
where $\mathcal{V}_{i}$ ($i=1,2,3$) are interaction vertices and $1/\mathcal{P}_j$ ($j=1,2$) and $1/\mathcal{P}_E$  correspond to the propagators. In Eq. (\ref{eqs:6S-nS-eta}), monopole form factor $\mathcal{F}(q^2,m_E) = (m_E^2 - \Lambda^2)/(q^2 - \Lambda^2)$ is adopted as suggested by a QCD sum rule study in Ref. \cite{Gortchakov:1995im} , in which we have introduced $m_E$ as the mass of the exchanged bottom meson and the cutoff $\Lambda = m_E + \alpha_\Lambda \Lambda_{QCD}$ with $\Lambda_{QCD}=220$ MeV \cite{Liu:2006dq,Liu:2009dr,Li:2013zcr}.
This included form factor is applied to mainly describe the structure effect of each vertex. In addition, this form factor also plays a role similar to the Pauli-Villas renormalization scheme which is often used to avoid ultraviolet divergence in the loop integrals \cite{Itzykson:1980rh,Peskin:1995ev}.

To describe the interaction vertices $\mathcal{V}_i$ in Eq. (\ref{eqs:6S-nS-eta}), in this paper the effective Lagrangian approach is adopted as has been done in Refs. \cite{Huang:2017kkg,Huang:2018cco}. With the heavy quark limit and chiral symmetry, the effective Lagrangians involved in concrete calculation are \cite{Casalbuoni:1996pg,Wang:2016qmz,Cheng:1992xi,Yan:1992gz,Wise:1992hn,Burdman:1992gh,Falk,Kaymakcalan:1983qq,Oh:2000qr,Colangelo:2002mj}
\begin{eqnarray}
\mathcal{L}_S&=& i g_1 \mathrm{Tr} \left[S^{(Q\bar{Q})} \bar{H}^{(\bar{Q}q)} \gamma^\mu \lrpartial_\mu \bar{H}^{(Q\bar{q})} \right] +H.c., \label{eqs:L-S}\\
\mathcal{L}_{\mathcal{P}} &=& i g_2 \mathrm{Tr} \left[ H_b^{(Q\bar{q})} \gamma_\mu \gamma_5 \mathcal{A}_{ba}^\mu \bar{H}_a^{(Q\bar{q})} \right]\label{eqs:L-G},
\end{eqnarray}
where ${S^{(Q\bar{Q})}}$ denotes the S-wave multiplets of the bottomonium \cite{Casalbuoni:1996pg}, ${H^{(Q\bar{q})}}$ represents the $H$ doublet ($\mathcal{B}$, $\mathcal{B}^*$)
\cite{Casalbuoni:1996pg,Kaymakcalan:1983qq,Oh:2000qr,Colangelo:2002mj}, and $\mathcal{A}_{ba}^\mu$ is the axial vector current of Nambu-Goldstone fields. The explicit expressions of these symbols are
\begin{eqnarray}
S^{(Q\bar{Q})}&=&\frac{1+\slashed{v}}{2}\Big[\Upsilon^\mu\gamma_\mu-\eta_b\gamma_5\Big]\frac{1-\slashed{v}}{2},\\
H^{(Q\bar{q})}&=&\frac{1+ \slashed{v}}{2} \left[\mathcal{B}^\ast_\mu \gamma^\mu-\mathcal{B} \gamma^5\right],\\
\mathcal{A}_\mu &=& \frac{1}{2} (\xi^\dag \partial_\mu \xi - \xi \partial_\mu \xi^\dag),
\end{eqnarray}
with $v^\mu$ being the 4-velocity, $\xi = e^{i\mathcal{M_P}/f_\pi}$. The pseudoscalar octet $\mathcal{M_P}$ reads as \cite{Chen:2012nva}
\begin{eqnarray}
\mathcal{M_P} &=&
 \left(
 \begin{array}{ccc}
\frac{\pi^0}{\sqrt{2}}+\alpha\eta+\beta\eta' & \pi^{+} & K^{+}\\
\pi^{-} & -\frac{\pi^0}{\sqrt{2}}+\alpha\eta+\beta\eta' &  K^{0}\\
 K^{-} & \bar{K}^{0} & \gamma\eta+\delta\eta'
 \end{array}
 \right),\label{eqs:octet}
\end{eqnarray}
where
\begin{equation}
\label{eqs:mixing}
 \begin{split}
\alpha&=\frac{\cos\theta-\sqrt{2}\sin\theta}{\sqrt{6}},\quad\quad\beta=\frac{\sin\theta+\sqrt{2}\cos\theta}{\sqrt{6}},\\
\gamma&=\frac{-2\cos\theta-\sqrt{2}\sin\theta}{\sqrt{6}},\quad\delta=\frac{-2\sin\theta+\sqrt{2}\cos\theta}{\sqrt{6}},
\end{split}
\end{equation}
and we take $\theta=-19.1^\circ$ \cite{Wang:2016qmz,Chen:2012nva,Coffman:1988ve,Jousset:1988ni}.

After expanding the Lagrangians in Eqs. (\ref{eqs:L-S}) and (\ref{eqs:L-G}), the interaction vertices $\mathcal{V}_i$ in Eq. (\ref{eqs:6S-nS-eta}) can be explicitly expressed by
\begin{eqnarray}
&&\mathcal{L}_{\Upsilon \mathcal{B}^{(\ast)} \mathcal{B}^{(\ast)}}\nonumber\\
&&= -ig_{\Upsilon \mathcal{BB} } \Upsilon_\mu (\partial^\mu
\mathcal{B} \mathcal{B}^\dagger- \mathcal{B}
\partial^\mu \mathcal{B}^\dagger) \nonumber\\ && \quad+ g_{\Upsilon
\mathcal{B}^\ast \mathcal{B}} \varepsilon^{\mu \nu \alpha \beta}
\partial_\mu \Upsilon_\nu (\mathcal{B}^\ast_\alpha \lrpartial_\beta
\mathcal{B}^\dagger -\mathcal{B} \lrpartial_\beta
\mathcal{B}_\alpha^{\ast \dagger} ) \nonumber\\ && \quad+ ig_{\Upsilon
\mathcal{B}^\ast \mathcal{B}^\ast} \Upsilon^\mu
(\mathcal{B}^\ast_\nu \partial^\nu \mathcal{B}^{\ast \dagger}_\mu
-\partial^\nu \mathcal{B}^{\ast}_\mu \mathcal{B}^{\ast \dagger}_\nu
-\mathcal{B}^\ast_\nu \lrpartial_\mu \mathcal{B}^{\ast \nu
\dagger}), \label{eqs:UpsilonBB}\\
&&\mathcal{L}_{\mathcal{B}^{(\ast)}\mathcal{B}^{(\ast)} \mathcal{M_P}}\nonumber\\
&&= i g_{\mathcal{B} \mathcal{B}^* \mathcal{M_P}} (\mathcal{B}^\dag \partial_\mu \mathcal{M_P} \mathcal{B}^{*\mu} - \mathcal{B}^{*\dag\mu} \partial_\mu \mathcal{M_P} \mathcal{B})\nonumber\\
&&\quad - g_{\mathcal{B}^* \mathcal{B}^* \mathcal{P}} \varepsilon^{\mu\nu\alpha\beta} \partial_\mu \mathcal{B}^{*\dag}_\nu \mathcal{M_P} \partial_\alpha \mathcal{B}^*_\beta,\label{eqs:HHP}
\end{eqnarray}
with 
$\mathcal{B}^{(\ast)}=(B^{(\ast)-},\bar{B}^{(\ast)0},\bar{B}_s^{(\ast)0})^T$.
The Feynman rules of all the vertices involved in our calculation are collected in the Appendix.

Finally, the decay widths of the processes $\Upsilon(6S) \to \Upsilon(nS) \eta^{(\prime)}$ can be evaluated by
\begin{eqnarray}
\Gamma = \frac{1}{24\pi} \frac{|\vec{p}_{\Upsilon(nS)}|}{m_{\Upsilon(6S)}^2}
|\overline{\mathcal{M}^{\mathrm{Total}}}|^2,
\end{eqnarray}
in which the average over polarizations of the initial $\Upsilon(6S)$ and the sum over those of the $\Upsilon(nS)$ have been done, and
\begin{eqnarray}\label{TDA}
\mathcal{M}^{\mathrm{Total}} = 4 \sum_{i=1}^{6} \mathcal{M}_i^q + 2 \sum_{i=1}^{6} \mathcal{M}_i^s.\label{eqs:M-total}
\end{eqnarray}
In Eq. (\ref{eqs:M-total}), $\sum_{i=1}^{6}$ means that all the diagrams shown in Fig. \ref{fig:6S-nS-eta-all} will give contributions to this loop-described transition $\Upsilon(6S) \to \Upsilon(nS) \eta^{(\prime)}$, $\mathcal{M}_i^q$ and $\mathcal{M}_i^s$ denote that in the amplitudes, the loops are composed of bottom ($B^{(\ast)}$) and bottom-strange ($B_s^{(\ast)}$) mesons, respectively, and the factors 4 and 2 come from the charge conjugation and the isospin transformations on the bridged $B_{(s)}^{(*)}$.

\section{Numerical results}\label{sec3}

Before presenting our numerical results, we want to illustrate how to determine the relevant coupling constants. For $g_{\Upsilon\mathcal{B}^{(*)}{\mathcal{B}}^{(*)}}$, we use the corresponding partial decay widths calculated via the potential model in Ref. \cite{Godfrey:2015dia} to fit them. The results are collected in Table \ref{tab:cc-6S-HH}.
\renewcommand{\arraystretch}{1.8}
\begin{table}[htpb]
\centering \caption{The partial decay widths of $\Upsilon(6S)\to \mathcal{B}^{(*)}\bar{\mathcal{B}}^{(*)}$ presented in Ref.~\cite{Godfrey:2015dia} and the extracted coupling constants \cite{Huang:2017kkg,Huang:2018cco}.\label{tab:cc-6S-HH}}
\begin{tabular}{ccccc}
\toprule[1pt]
Final state  &~& Decay width (MeV) &~& Coupling constant\\
\midrule[0.6pt] %
$B \bar{B}$ &~& 1.32 &~& 0.654\\
$B \bar{B}^{*}$ &~& 7.59 &~& $0.077~\mathrm{GeV}^{-1}$\\
$B^* \bar{B}^*$ &~& 5.89 &~& 0.611\\
$B_s \bar{B}_s$ &~& $1.31\times10^{-3}$ &~& 0.043\\
$B_s \bar{B}_s^{*}$ &~& 0.136 &~& $0.023~\mathrm{GeV}^{-1}$\\
$B_s^* \bar{B}_s^*$ &~& 0.310 &~& 0.354\\
\bottomrule[1pt]
\end{tabular}
\end{table}

For $g_{\Upsilon\mathcal{B}^{(*)}{\mathcal{B}}^{(*)}}$, we need to consider the internal symmetry implied by the heavy quark effective theory, under which all the coupling constants are related to each other through a global constant $g_1$ in Eq. (\ref{eqs:L-S}). With the help of the vector meson dominance ansatz \cite{Achasov:1994vh,Deandrea:2003pv,Colangelo:2003sa,Lin:1999ad}, a relation among these couplings can be given as
\begin{eqnarray}
g_{\Upsilon \mathcal{BB}}=g_{\Upsilon \mathcal{B}^\ast \mathcal{B}^\ast} \frac{m_\mathcal{B}}{m_{\mathcal{B}^\ast}}=g_{\Upsilon \mathcal{B}^\ast \mathcal{B}} \sqrt{m_\mathcal{B} m_{\mathcal{B}^\ast}}=\frac{m_{\Upsilon}}{f_{\Upsilon}},
\end{eqnarray}
with $m_{\Upsilon}$ and $f_{\Upsilon}$ being the mass and decay constant of $\Upsilon$ \cite{Achasov:1994vh,Deandrea:2003pv,Colangelo:2003sa,Lin:1999ad}. The decay constant of $\Upsilon(nS)$, i.e., $f_\Upsilon$, can be obtained by fitting to the leptonic decay width $\Gamma[\Upsilon(nS)\to e^+e^-]$ given in Ref. \cite{Olive:2016xmw} by
\begin{eqnarray}
\Gamma_{V\to e^+e^-}=\frac{4\pi}{3}\frac{\alpha^2}{M_V}f_V^2C_V,
\end{eqnarray}
where $\alpha$ is the fine-structure constant and $C_V=1/9$ for the $\Upsilon(nS)$ meson. We estimate the different values of $g_1$ and list them in Table \ref{tab:nS-decay-constants}.
\begin{table}[htpb]
\centering \caption{The values of the coupling constant $g_1$ in Eq.~(\ref{eqs:L-S}). Here, we also list the experimental
	leptonic decay widths \cite{Olive:2016xmw} and the extracted decay constants.\label{tab:nS-decay-constants}}
 \renewcommand{\arraystretch}{1.8}
\begin{tabular}{ccccccc}
\toprule[1pt]
Particle  &~& Decay width (MeV)&~& Decay constant (GeV)&~& $g_1$ GeV$^{-3/2}$\\
\midrule[0.6pt] %
$\Upsilon(1S)$ &~& $1.34\times10^{-3}$ &~& 0.715 &~& 0.407 \\
$\Upsilon(2S)$ &~& $0.612\times10^{-3}$&~& 0.497&~& 0.603\\
$\Upsilon(3S)$ &~& $0.443\times10^{-3}$ &~& 0.430&~& 0.709\\
\bottomrule[1pt]
\end{tabular}
\end{table}

Similarly, $g_{\mathcal{B}^{(*)}{\mathcal{B}}^{(*)}\eta^{(\prime)}}$ can also be connected by a coupling constant $g_2=g_\pi/f_\pi$ with $f_\pi$=131 MeV and $g_\pi$=0.569 shown in Eq. (\ref{eqs:L-G}), i.e.,
\begin{eqnarray}
\frac{g_{BB^*\eta}}{\sqrt{m_B m_{B^*}}} &=& g_{B^* B^* \eta} = 2 \alpha \frac{g_\pi}{f_\pi}\nonumber,\\
\frac{g_{B_sB_s^*\eta}}{\sqrt{m_{B_s} m_{B_s^*}}} &=& g_{B_s^* B_s^* \eta} = 2 \gamma \frac{g_\pi}{f_\pi}\nonumber,\\
\frac{g_{BB^*\eta'}}{\sqrt{m_B m_{B^*}}} &=& g_{B^* B^* \eta'} = 2 \beta \frac{g_\pi}{f_\pi}\nonumber,\\
\frac{g_{B_sB_s^*\eta'}}{\sqrt{m_{B_s} m_{B_s^*}}} &=& g_{B_s^* B_s^* \eta'} = 2 \delta \frac{g_\pi}{f_\pi}\nonumber,
\end{eqnarray}
in which $\alpha,~\beta,~\gamma,~\delta$ have been defined in Eq. (\ref{eqs:mixing}).

Besides these fixed coupling constants, there is still a free parameter $\alpha_\Lambda$, which is included in the form factor $\mathcal{F}(q^2,m_E)$ in the amplitudes. According to the experiences in Refs. \cite{Meng:2007tk,Meng:2008dd,Meng:2008bq}, $\alpha_\Lambda=3$ is taken in our calculation.

Our calculation gives the partial decay widths for the processes $\Upsilon(6S)\to\Upsilon(nS)\eta^{(\prime)}$ with $\alpha_\Lambda=3$ as
\begin{equation}\label{eqs:width}
\begin{split}
\Gamma[\Upsilon(6S) \to \Upsilon(1S) \eta] &= 197.518~\mathrm{keV},\\
\Gamma[\Upsilon(6S) \to \Upsilon(2S) \eta] &= 118.762~\mathrm{keV},\\
\Gamma[\Upsilon(6S) \to \Upsilon(3S) \eta] &= 8.264~\mathrm{keV},\\
\Gamma[\Upsilon(6S) \to \Upsilon(1S) \eta'] &= 88.347~\mathrm{keV},\\
\Gamma[\Upsilon(6S) \to \Upsilon(2S) \eta'] &= 0.080~\mathrm{keV}.
\end{split}
\end{equation}
The corresponding branching ratios are extracted as
\begin{equation}\label{eqs:ratio}
\begin{split}
\mathcal{B}[\Upsilon(6S) \to \Upsilon(1S) \eta] &= 3.238 \times 10^{-3} , \\
\mathcal{B}[\Upsilon(6S) \to \Upsilon(2S) \eta] &= 1.947 \times 10^{-3} , \\
\mathcal{B}[\Upsilon(6S) \to \Upsilon(3S) \eta] &= 0.135 \times 10^{-3}, \\
\mathcal{B}[\Upsilon(6S) \to \Upsilon(1S) \eta'] &= 1.448 \times 10^{-3}, \\
\mathcal{B}[\Upsilon(6S) \to \Upsilon(2S) \eta'] &= 1.305 \times 10^{-6}.
\end{split}
\end{equation}

The above results indicate that most of the hidden-bottom $\eta^{(\prime)}$ transitions of the $\Upsilon(6S)$ have branching ratios around $10^{-3}$. Because of these predicted considerable branching ratios, we are confident that all these transitions can be observed in future experiments like Belle II. For the process $\Upsilon(6S) \to \Upsilon(2S) \eta^{(\prime)}$, it is easy to see that although its branching ratio is 3 orders of magnitude lower than others, we note here that this smallness is mainly due to the restriction of phase space together with the $\eta-\eta'$ mixing depicted in Eqs. (\ref{eqs:octet}) and (\ref{eqs:mixing}), which gives opposite contributions to $\mathcal{M}_i^q$ and $\mathcal{M}_i^s$ in Eq. (\ref{eqs:M-total}). Although this also happens in $\Upsilon(6S)\to\Upsilon(1S)\eta^{(\prime)}$, its branching ratio can still hold at $10^{-3}$ due to the larger phase space.

We also find that the ratios of these calculated branching ratios are insensitive to the $\alpha_\Lambda$ values. When scanning $\alpha_\Lambda$ in a range $[2.7,3.3]$, we present the ratios, which reflects this insensitivity well. In the following equations, we list the typical ratios
\begin{equation}\label{ratios}
\begin{split}
\frac{\mathcal{B}[\Upsilon(6S) \to \Upsilon(2S) \eta]}{\mathcal{B}[\Upsilon(6S) \to \Upsilon(1S) \eta]} &= 0.579 \sim 0.626, \\
\frac{\mathcal{B}[\Upsilon(6S) \to \Upsilon(3S) \eta]}{\mathcal{B}[\Upsilon(6S) \to \Upsilon(1S) \eta]} &= 0.039 \sim 0.045, \\
\frac{\mathcal{B}[\Upsilon(6S) \to \Upsilon(3S) \eta]}{\mathcal{B}[\Upsilon(6S) \to \Upsilon(2S) \eta]} &= 0.067 \sim 0.072, \\
\frac{\mathcal{B}[\Upsilon(6S) \to \Upsilon(2S) \eta']}{\mathcal{B}[\Upsilon(6S) \to \Upsilon(1S) \eta']} &= (8.739\sim 9.305) \times 10^{-4},
\end{split}
\end{equation}
and
\begin{equation}
\begin{split}
\frac{\mathcal{B}[\Upsilon(6S) \to \Upsilon(1S) \eta]}{\mathcal{B}[\Upsilon(6S) \to \Upsilon(1S) \eta']} &= 2.231 \sim 2.241, \\
\frac{\mathcal{B}[\Upsilon(6S) \to \Upsilon(1S) \eta]}{\mathcal{B}[\Upsilon(6S) \to \Upsilon(2S) \eta']} &= (2.408 \sim 2.553) \times 10^3, \\
\frac{\mathcal{B}[\Upsilon(6S) \to \Upsilon(2S) \eta]}{\mathcal{B}[\Upsilon(6S) \to \Upsilon(1S) \eta']} &= 1.292 \sim 1.403, \\
\frac{\mathcal{B}[\Upsilon(6S) \to \Upsilon(2S) \eta]}{\mathcal{B}[\Upsilon(6S) \to \Upsilon(2S) \eta']} &= (1.478 \sim 1.508) \times 10^3, \\
\frac{\mathcal{B}[\Upsilon(6S) \to \Upsilon(3S) \eta]}{\mathcal{B}[\Upsilon(6S) \to \Upsilon(1S) \eta']} &= 0.087 \sim 0.101, \\
\frac{\mathcal{B}[\Upsilon(6S) \to \Upsilon(3S) \eta]}{\mathcal{B}[\Upsilon(6S) \to \Upsilon(2S) \eta']} &= (0.995 \sim 1.089) \times 10^2,
\end{split}
\end{equation}
which, in our view, can be tested by a future experiment.

\section{Conclusions and discussion}

Two very recent experimental observations of the hadronic transitions $\Upsilon(5S) \to \Upsilon(1^3D_J)\eta$ \cite{Tamponi:2018cuf} and $\Upsilon(6S)\to \chi_{bJ}\pi^+\pi^-\pi^0$ \cite{Yin:2018ojs} again enforced our ambition to explore the hidden-bottom decays of higher bottomonia, since these Belle measurements are consistent with our predictions made in Refs. \cite{Wang:2016qmz,Huang:2017kkg}. Here, the hadronic loop mechanism may play a crucial role in mediating these hidden-bottom transitions of higher bottomonia to the final states.

In this work, we have applied the hadronic loop mechanism to investigate the $\Upsilon(6S)$ decays into $\Upsilon(nS)$ plus $\eta$ or $\eta^\prime$, which are still missing in an experiment. Just illustrated in the former section, these discussed transitions have considerable branching ratios, which make the experimental exploration of them accessible, especially for the Belle II experiment.

At present, the knowledge of the $\Upsilon(11020)$ is still not enough due to the absence of the relevant experimental information \cite{Olive:2016xmw}. Since the present work is one part of the whole study on the $\Upsilon(11020)$, we hope that these predictions given in this paper can stimulate the experimentalist's interest in carrying out the search for the discussed transitions. This will make the information of the $\Upsilon(11020)$ more and more abundant. Surely, if these predicted branching ratios and partial widths can be confirmed in a future experiment, the importance of the hadronic loop mechanism to the hidden-bottom transitions of higher bottomonia can be further realized. To some extent, it is also helpful to deepen our understanding of the nonperturbative behavior of strong interaction.

\section*{Acknowledgments}

This project is supported by the National Natural Science Foundation of China under Grants No.~11222547 and No. 11175073.
Xiang Liu is also supported by the National Program for Support of Young Top-Notch Professionals.

\section*{Appendix}\label{appendixA}
The Feynman rules for each interaction vertex involved in our calculation are

\begin{eqnarray}
\raisebox{-15pt}{\includegraphics[width=0.16%
\textwidth]{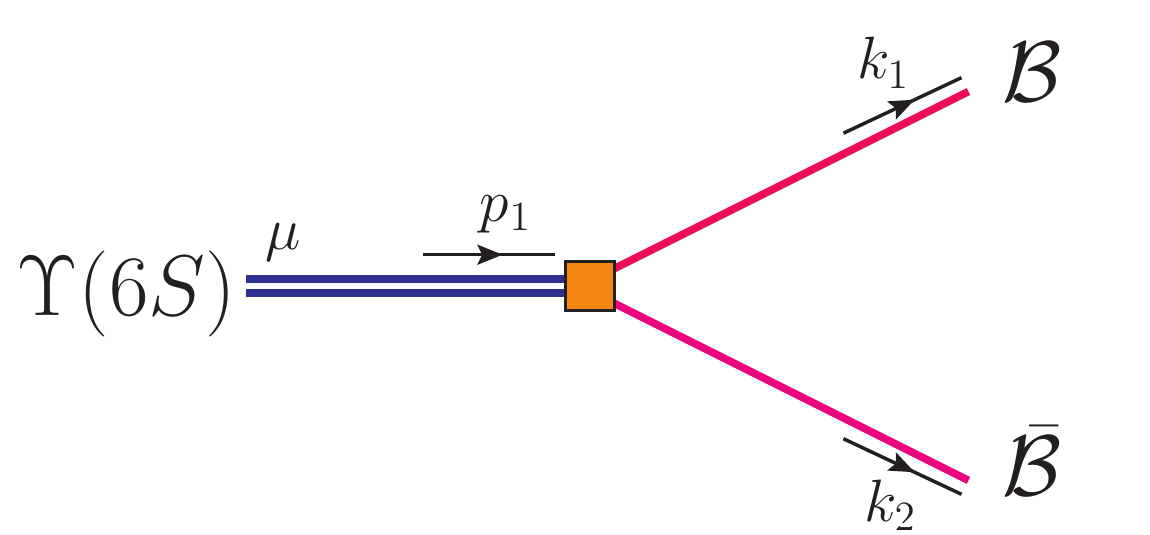}} &\widehat{=}& g_{\Upsilon' \mathcal{B} \bar{\mathcal{B}}} \epsilon^\mu_{\Upsilon'} (k_{1\mu}-k_{2\mu}),\\
\raisebox{-15pt}{\includegraphics[width=0.16%
\textwidth]{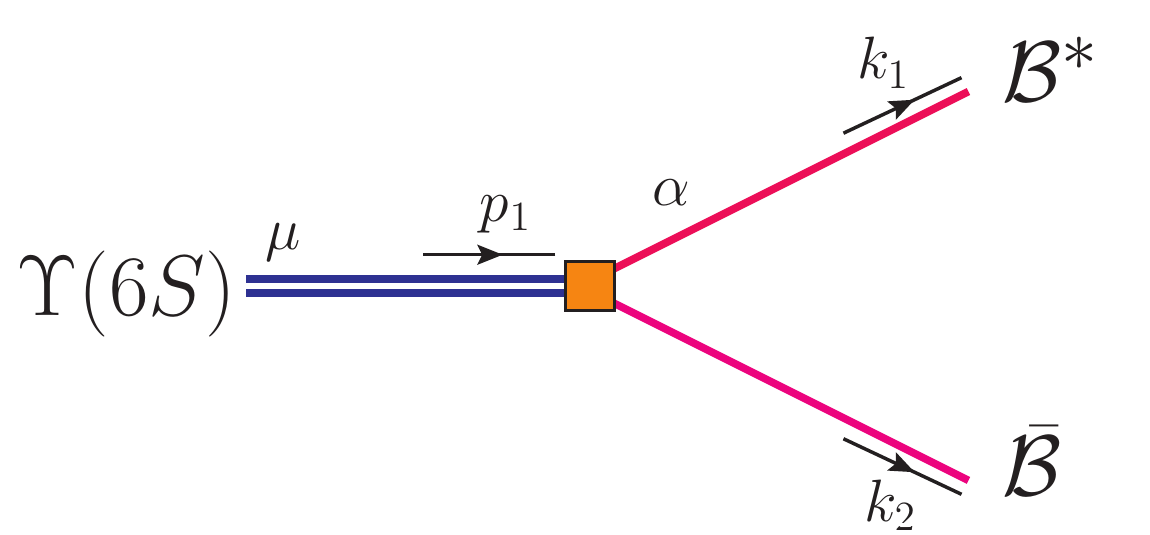}} &\widehat{=}& g_{\Upsilon' \mathcal{B}^* \bar{\mathcal{B}}} \varepsilon^{\mu\nu\alpha\beta} \epsilon_{\Upsilon' \mu} p_{1\nu} (k_{1\beta}-k_{2\beta}),\\
\raisebox{-15pt}{\includegraphics[width=0.16%
\textwidth]{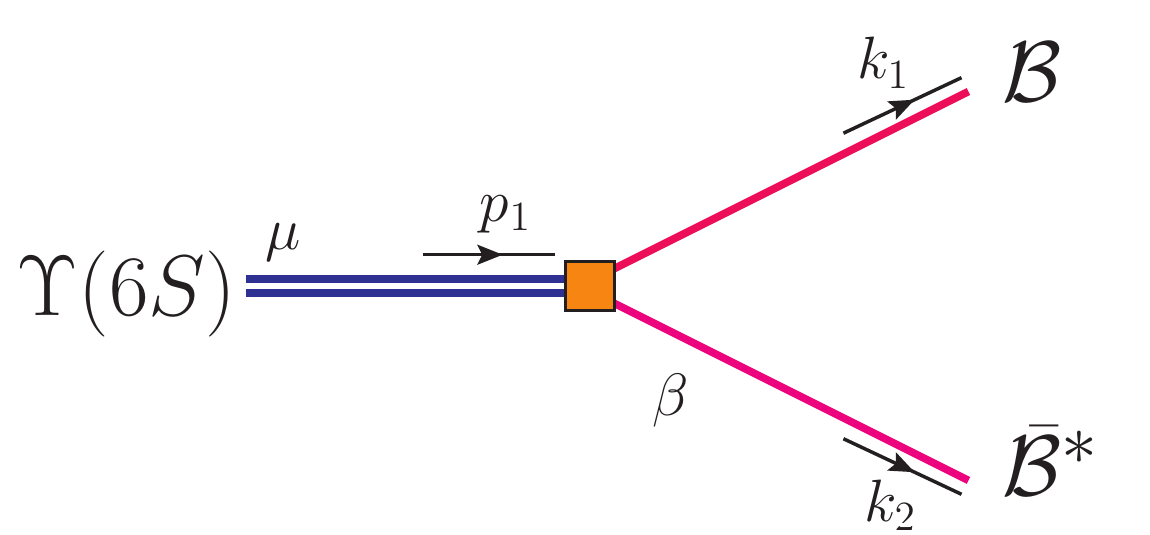}} &\widehat{=}& g_{\Upsilon' \mathcal{B} \bar{\mathcal{B}}^*} \varepsilon^{\mu\nu\alpha\beta} \epsilon_{\Upsilon'\mu} p_{1\nu} (k_{1\beta} - k_{2\beta}) ,\\
\raisebox{-15pt}{\includegraphics[width=0.16%
\textwidth]{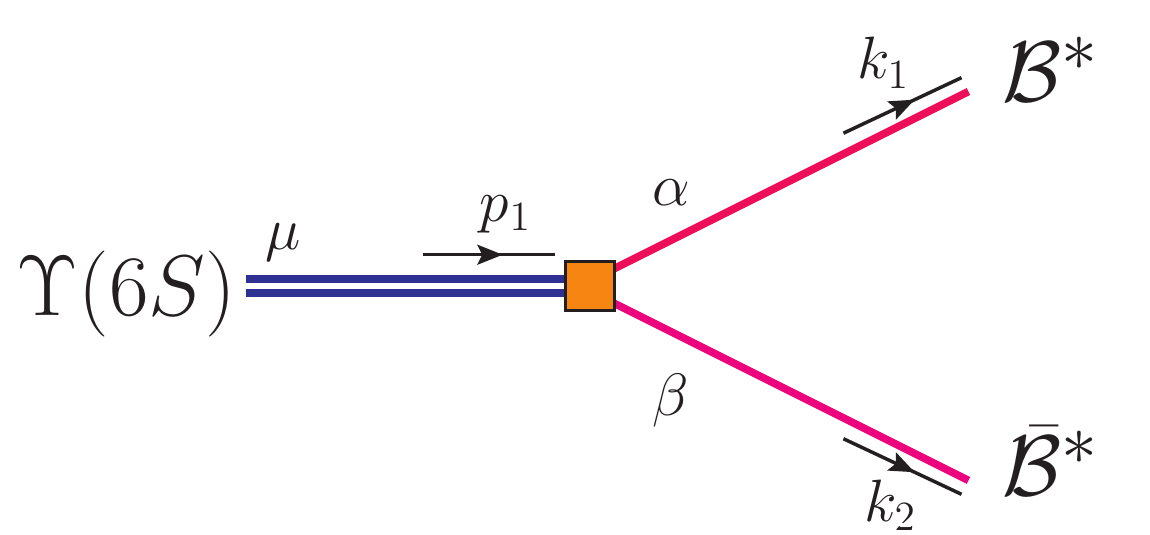}} &\widehat{=}& g_{\Upsilon' \mathcal{B}^* \bar{\mathcal{B}}^*} \epsilon_{\Upsilon'}^\mu \left(g_{\mu\alpha}k_{2\beta} - g_{\mu\beta}k_{1\alpha}  \right.\nonumber\\
&&\left. + g_{\alpha\beta}(k_{1\mu}-k_{2\mu})\right),
\end{eqnarray}

\begin{eqnarray}
\raisebox{-15pt}{\includegraphics[width=0.16%
\textwidth]{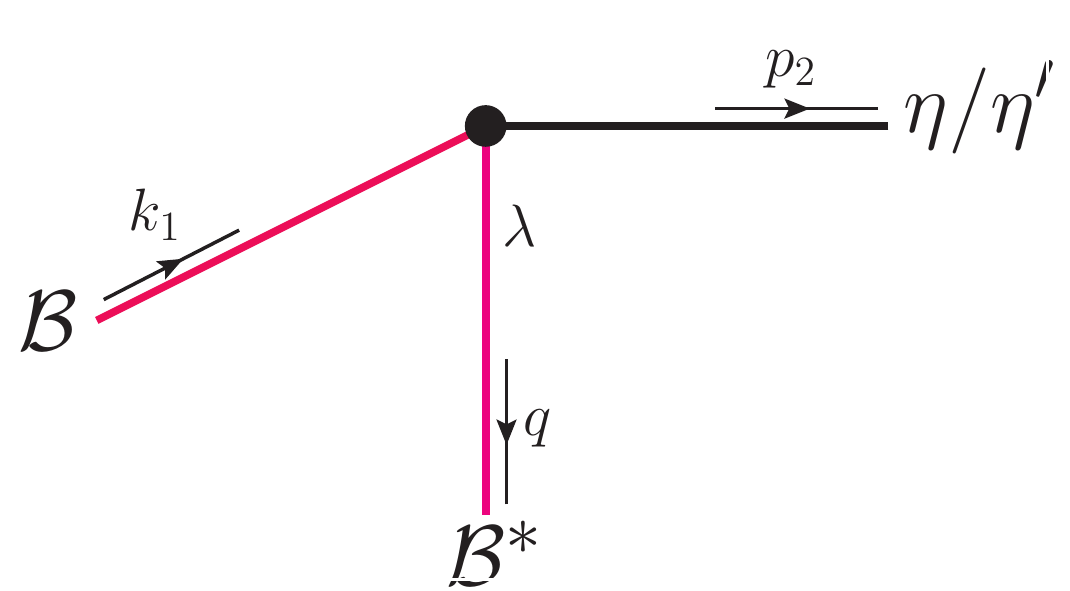}} &\widehat{=}& g_{\mathcal{B} \mathcal{B}^* \tilde{\eta}} p_{2\lambda},\\
\raisebox{-15pt}{\includegraphics[width=0.16%
\textwidth]{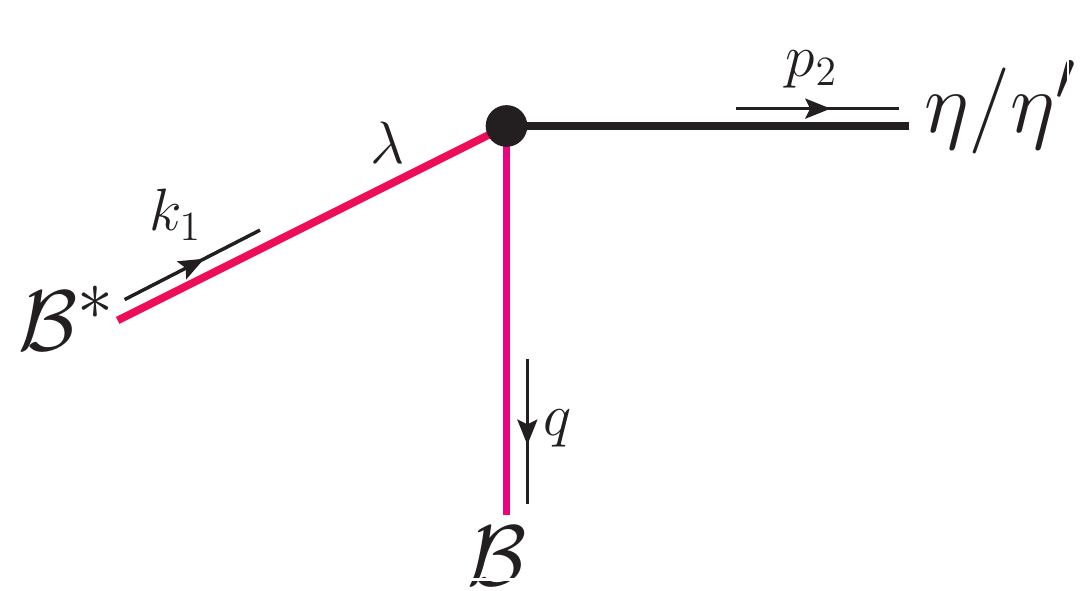}} &\widehat{=}& g_{\mathcal{B}^* \mathcal{B} \tilde{\eta}} p_{2\lambda} ,\\
\raisebox{-15pt}{\includegraphics[width=0.16%
\textwidth]{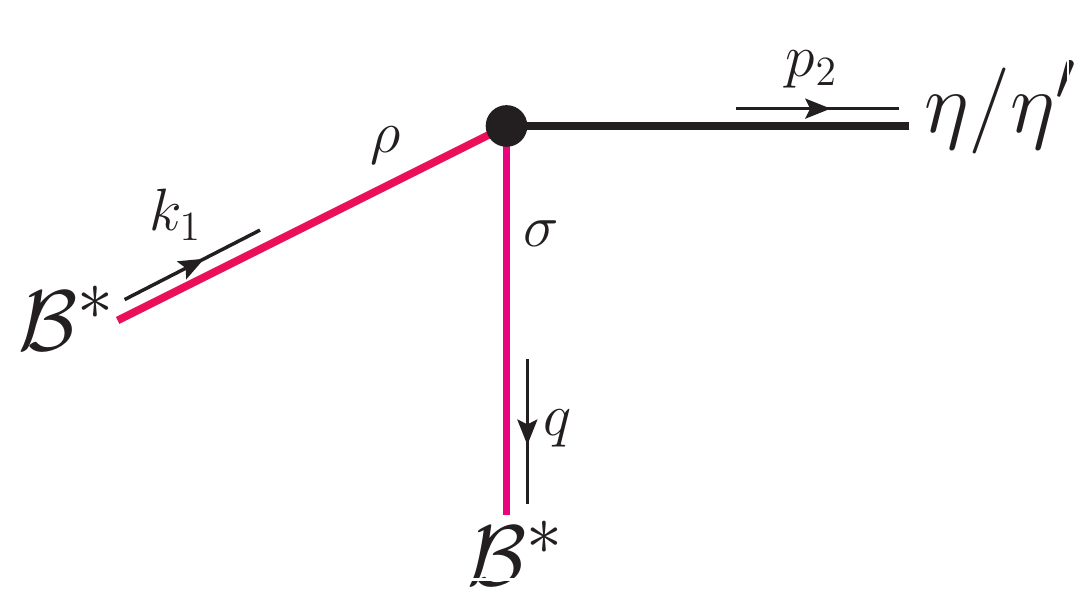}} &\widehat{=}& g_{\mathcal{B}^* \mathcal{B}^* \tilde{\eta}} \varepsilon^{\lambda\rho\delta\sigma} k_{1\lambda} q_\delta,
\end{eqnarray}
and
\begin{eqnarray}
\raisebox{-15pt}{\includegraphics[width=0.16%
\textwidth]{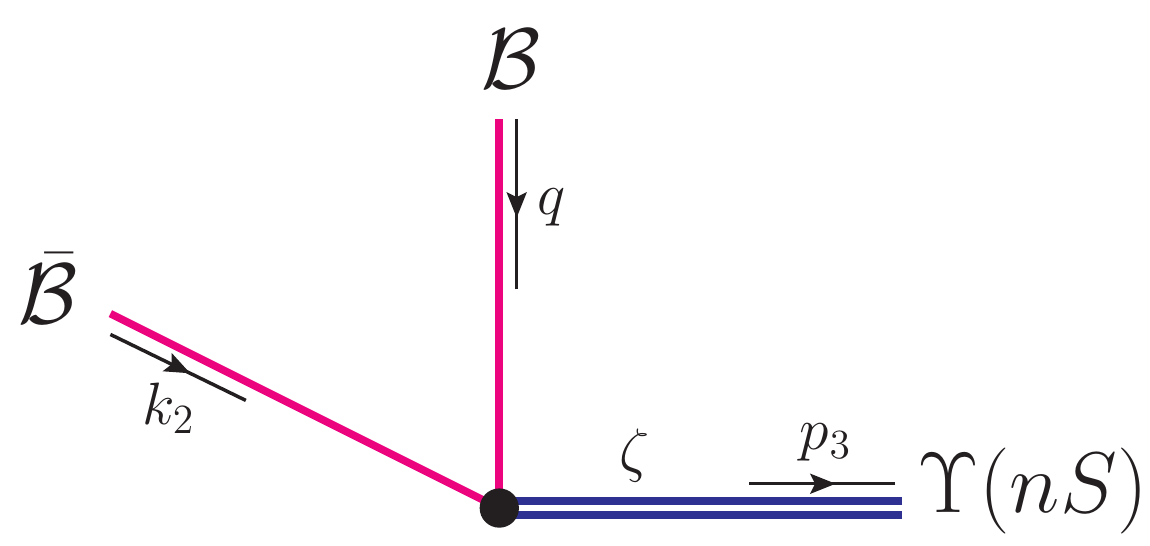}} &\widehat{=}& g_{\bar{\mathcal{B}} \mathcal{B} \Upsilon} \epsilon^{*\zeta}_{\Upsilon}(q_\zeta-k_{2\zeta}),\\
\raisebox{-15pt}{\includegraphics[width=0.16%
\textwidth]{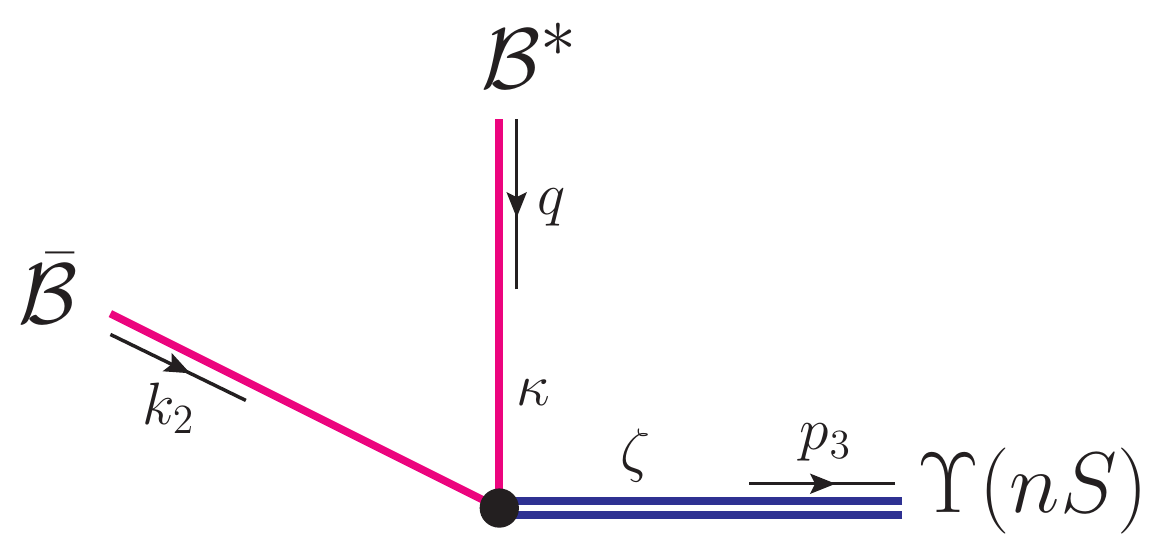}} &\widehat{=}& g_{\bar{\mathcal{B}} \mathcal{B}^* \Upsilon} \varepsilon^{\zeta\eta\kappa\xi} \epsilon^*_{\Upsilon\zeta} p_{3\eta} (q_\xi-k_{2\xi}),\\
\raisebox{-15pt}{\includegraphics[width=0.16%
\textwidth]{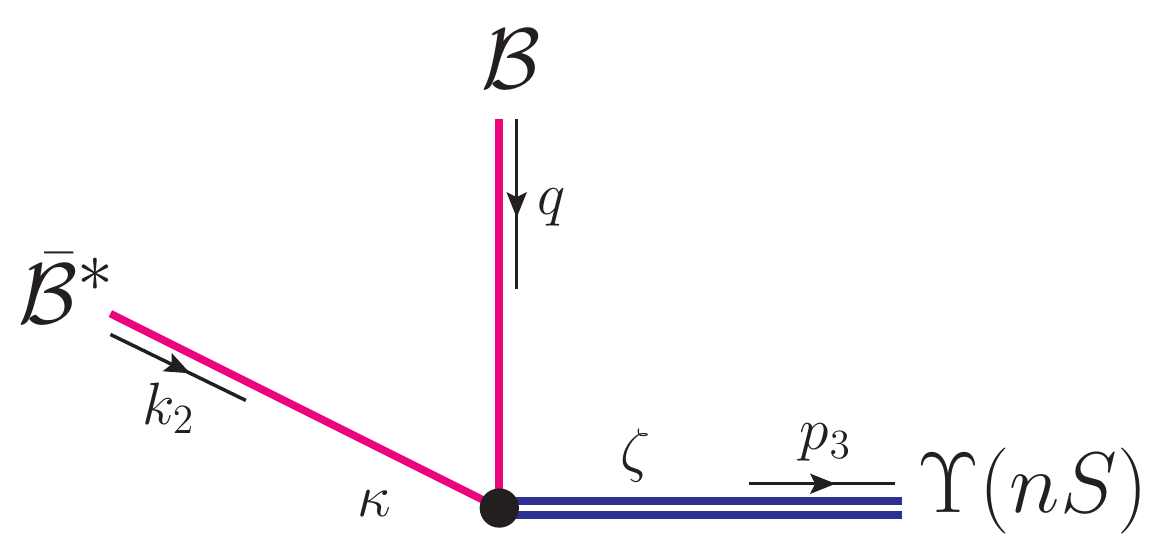}} &\widehat{=}& g_{\bar{\mathcal{B}}^* \mathcal{B} \Upsilon} \varepsilon^{\zeta\eta\kappa\xi} \epsilon^*_{\Upsilon\zeta} p_{3\eta} (k_{2\xi}-q_\xi),\\
\raisebox{-15pt}{\includegraphics[width=0.16%
\textwidth]{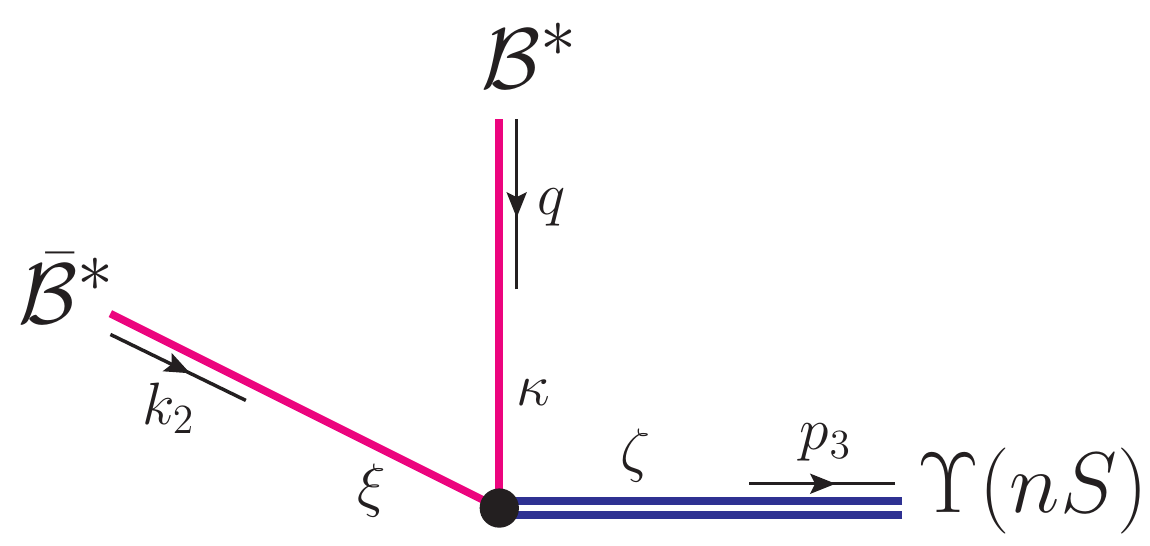}} &\widehat{=}& g_{\bar{\mathcal{B}}^* \mathcal{B}^* \Upsilon} \epsilon^{*\zeta}_{\Upsilon}\left(g_{\zeta\xi} q_\kappa - g_{\zeta\kappa} k_{2\xi} \right. \nonumber\\
&&\left.  - g_{\kappa\xi} (q_\zeta-k_{2\zeta})\right).
\end{eqnarray}

In addition, the involved propagators are given by
\begin{eqnarray}
\raisebox{0pt}{\includegraphics[width=0.16%
\textwidth]{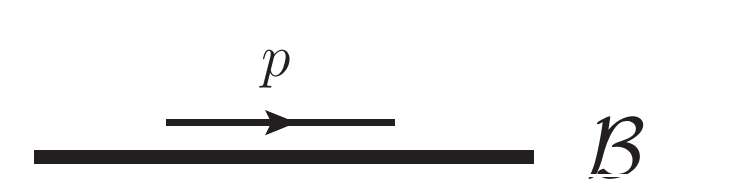}} &\widehat{=}& \frac{i}{p^2-m_\mathcal{B}^2},\\
\raisebox{0pt}{\includegraphics[width=0.16%
\textwidth]{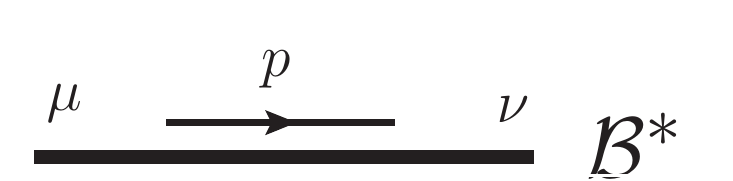}} &\widehat{=}& \frac{i(-g^{\mu \nu} +p^\mu p^\nu/m_{\mathcal{B}^*}^2)}{p^2-m_{\mathcal{B}^*}^2}.
\end{eqnarray}

\vfil


\begin{thebibliography}{}

\bibitem{Abe:2007tk}
  K.~F.~Chen {\it et al.}  [Belle Collaboration],
  Observation of anomalous $\Upsilon(1S) \pi^+ \pi^-$ and $\Upsilon(2S) \pi^+ \pi^-$ production near the $\Upsilon(5S)$ resonance,
  Phys.\ Rev.\ Lett.\  {\bf 100}, 112001 (2008).

\bibitem{He:2014sqj}
  X.~H.~He {\it et al.} [Belle Collaboration],
  Observation of $e^+e^- \to \pi^+ \pi^- \pi^0 \chi_{bJ}$ and Search for $X_b \to \omega \Upsilon(1S)$ at $\sqrt{s}=10.867$ GeV,
  Phys.\ Rev.\ Lett.\  {\bf 113}, 142001 (2014).

\bibitem{Belle:2011aa}
  A.~Bondar {\it et al.} [Belle Collaboration],
  Observation of two charged bottomonium-like resonances in $\Upsilon(5S)$ decays,
  Phys.\ Rev.\ Lett.\  {\bf 108}, 122001 (2012).


\bibitem{Krokovny:2013}
 P.~Krokovny {\it et al.}  [Belle Collaboration],
  Phys.\ Rev.\ D {\bf 88}, 052016 (2013).

\bibitem{Olive:2016xmw}
  C.~Patrignani {\it et al.} [Particle Data Group Collaboration],
  Review of Particle Physics,
  Chin.\ Phys.\ C {\bf 40}, 100001 (2016).





\bibitem{Chen:2011zv}
  D.~Y.~Chen, X.~Liu and S.~L.~Zhu,
  Charged bottomonium-like states $Z_b(10610)$ and $Z_b(10650)$ and the $\Upsilon(5S)\to \Upsilon(2S)\pi^+\pi^-$ decay,
  Phys.\ Rev.\ D {\bf 84}, 074016 (2011).

\bibitem{Meng:2007tk}
  C.~Meng and K.~T.~Chao,
  Scalar resonance contributions to the dipion transition rates of $\Upsilon(4S,5S)$ in the re-scattering model,
  Phys.\ Rev.\ D {\bf 77}, 074003 (2008).

\bibitem{Meng:2008dd}
  C.~Meng and K.~T.~Chao,
  Peak shifts due to $B^\ast\bar{B}^\ast$ rescattering in $\Upsilon(5S)$ dipion transitions,
  Phys.\ Rev.\ D {\bf 78}, 034022 (2008).

\bibitem{Meng:2008bq}
  C.~Meng and K.~T.~Chao,
  $\Upsilon(4S,5S)\to \Upsilon(1S)\eta$ transitions in the rescattering model and the new BaBar measurement,
  Phys.\ Rev.\ D {\bf 78}, 074001 (2008).

\bibitem{Simonov:2008qy}
  Y.~A.~Simonov and A.~I.~Veselov,
  Bottomonium dipion transitions,
  Phys.\ Rev.\ D {\bf 79}, 034024 (2009).

\bibitem{Chen:2011qx}
  D.~Y.~Chen, J.~He, X.~Q.~Li and X.~Liu,
  Dipion invariant mass distribution of the anomalous $\Upsilon(1S) \pi^{+} \pi^{-}$ and $\Upsilon(2S) \pi^{+} \pi^{-}$ production near the peak of $\Upsilon(10860)$,
  Phys.\ Rev.\ D {\bf 84}, 074006 (2011).

\bibitem{Chen:2014ccr}
  D.~Y.~Chen, X.~Liu and T.~Matsuki,
  Explaining the anomalous $\Upsilon(5S)\to \chi_{bJ}\omega$ decays through the hadronic loop effect,
  Phys.\ Rev.\ D {\bf 90}, 034019 (2014).

\bibitem{Chen:2011pv}
  D.~Y.~Chen and X.~Liu,
  $Z_b(10610)$ and $Z_b(10650)$ structures produced by the initial single pion emission in the $\Upsilon(5S)$ decays,
  Phys.\ Rev.\ D {\bf 84}, 094003 (2011).

\bibitem{Wang:2016qmz}
  B.~Wang, X.~Liu and D.~Y.~Chen,
  Phys.\ Rev.\ D {\bf 94}, no. 9, 094039 (2016)
  [arXiv:1611.02369 [hep-ph]].

\bibitem{Tamponi:2018cuf}
  U.~Tamponi {\it et al.} [Belle Collaboration],
  Inclusive study of bottomonium production in association with an $\eta$ meson in $e^+e^-$ annihilations near $\Upsilon(5S)$,
  [arXiv:1803.03225 [hep-ex]].

\bibitem{Yin:2018ojs}
  J.~H.~Yin {\it et al.} [Belle Collaboration],
  Observation of $e^+e^-\to\pi^+\pi^-\pi^0\chi_{b1,2}(1P)$ and search for $e^+e^-\to\phi\chi_{b1,2}(1P)$ at $\sqrt{s}=10.96-11.05$ GeV,
  arXiv:1806.06203 [hep-ex].


  
\bibitem{Huang:2017kkg} 
  Q.~Huang, B.~Wang, X.~Liu, D.~Y.~Chen and T.~Matsuki,
  Exploring the $\Upsilon (6S)\rightarrow \chi _{bJ}\phi $ and $\Upsilon (6S)\rightarrow \chi _{bJ}\omega $ hidden-bottom hadronic transitions,
  Eur.\ Phys.\ J.\ C {\bf 77}, no. 3, 165 (2017)
  [arXiv:1701.00894 [hep-ph]].
  
  

\bibitem{Huang:2018cco}
  Q.~Huang, H.~Xu, X.~Liu and T.~Matsuki,
  Potential observation of the $\Upsilon(6S) \to \Upsilon(1^3D_J) \eta$ transitions at Belle II,
  Phys.\ Rev.\ D {\bf 97}, no. 9, 094018 (2018)
  [arXiv:1804.01017 [hep-ph]].

\bibitem{Gortchakov:1995im}
  O.~Gortchakov, M.~P.~Locher, V.~E.~Markushin and S.~von Rotz,
  Two meson doorway calculation for $\bar{p} p \to \phi \pi$ including off-shell effects and the OZI rule,
  Z.\ Phys.\ A {\bf 353}, 447 (1996).

\bibitem{Liu:2006dq}
X.~Liu, X.~Q.~Zeng and X.~Q.~Li,
Study on contributions of hadronic loops to decays of $J/\psi\to$ vector + pseudoscalar mesons,
Phys.\ Rev.\ D {\bf 74}, 074003 (2006)

\bibitem{Liu:2009dr}
  X.~Liu, B.~Zhang and X.~Q.~Li,
  The Puzzle of excessive non-$D\bar{D}$ component of the inclusive $\psi(3770)$ decay and the long-distant contribution,
  Phys.\ Lett.\ B {\bf 675}, 441 (2009).

\bibitem{Li:2013zcr}
  G.~Li, X.~H.~Liu, Q.~Wang and Q.~Zhao,
  Further understanding of the non-$D\bar{D}$ decays of $\psi(3770)$,
  Phys.\ Rev.\ D {\bf 88}, 014010 (2013).


\bibitem{Itzykson:1980rh}
  C.~Itzykson and J.~B.~Zuber,
  Quantum Field Theory, New York, Usa: Mcgraw-hill (1980) 705 p. 

\bibitem{Peskin:1995ev}
  M.~E.~Peskin and D.~V.~Schroeder,
  ``An Introduction to quantum field theory,''  Reading, USA: Addison-Wesley (1995) 842 p.

\bibitem{Casalbuoni:1996pg}
  R.~Casalbuoni, A.~Deandrea, N.~Di Bartolomeo, R.~Gatto, F.~Feruglio and G.~Nardulli,
  Phenomenology of heavy meson chiral Lagrangians,
  Phys.\ Rept.\  {\bf 281}, 145 (1997).

\bibitem{Cheng:1992xi}
  H.~Y.~Cheng, C.~Y.~Cheung, G.~L.~Lin, Y.~C.~Lin, T.~M.~Yan and H.~L.~Yu,
  Chiral Lagrangians for radiative decays of heavy hadrons,
  Phys.\ Rev.\  D {\bf 47}, 1030 (1993).

\bibitem{Yan:1992gz}
  T.~M.~Yan, H.~Y.~Cheng, C.~Y.~Cheung, G.~L.~Lin, Y.~C.~Lin and H.~L.~Yu,
  Heavy Quark Symmetry And Chiral Dynamics,
  Phys.\ Rev.\  D {\bf 46}, 1148 (1992).

\bibitem{Wise:1992hn}
  M.~B.~Wise,
  Chiral Perturbation Theory For Hadrons Containing A Heavy Quark,
  Phys.\ Rev.\  D {\bf 45}, R2188 (1992).

\bibitem{Burdman:1992gh}
  G.~Burdman and J.~F.~Donoghue,
  Union of chiral and heavy quark symmetries,
  Phys.\ Lett.\  B {\bf 280}, 287 (1992).

\bibitem{Falk}
  A.~F.~Falk and M.~E.~Luke,
  Strong decays of excited heavy mesons in chiral perturbation theory,
  Phys.\ Lett.\ B {\bf 292}, 119 (1992).

\bibitem{Kaymakcalan:1983qq}
  O.~Kaymakcalan, S.~Rajeev and J.~Schechter,
  Nonabelian Anomaly and Vector Meson Decays,
  Phys.\ Rev.\ D {\bf 30}, 594 (1984).

\bibitem{Oh:2000qr}
  Y.~s.~Oh, T.~Song and S.~H.~Lee,
  $J/\psi$ absorption by $\pi$ and $\rho$ mesons in meson exchange model with anomalous parity interactions,
  Phys.\ Rev.\ C {\bf 63}, 034901 (2001).

\bibitem{Colangelo:2002mj}
  P.~Colangelo, F.~De Fazio and T.~N.~Pham,
  $B \to K^- \chi_{c0}$ decay from charmed meson rescattering,
  Phys.\ Lett.\ B {\bf 542}, 71 (2002).

\bibitem{Chen:2012nva}
  D.~Y.~Chen, X.~Liu and T.~Matsuki,
  $\eta$ transitions between charmonia with meson loop contributions,
  Phys.\ Rev.\ D {\bf 87}, no. 5, 054006 (2013)
  [arXiv:1209.0064 [hep-ph]].

\bibitem{Coffman:1988ve}
  D.~Coffman {\it et al.} [MARK-III Collaboration],
  Measurements of $J/\psi$ Decays Into a Vector and a Pseudoscalar Meson,
  Phys.\ Rev.\ D {\bf 38}, 2695 (1988)
  Erratum: [Phys.\ Rev.\ D {\bf 40}, 3788 (1989)].

\bibitem{Jousset:1988ni}
  J.~Jousset {\it et al.} [DM2 Collaboration],
  The $J/\psi \to$ Vector+Pseudoscalar Decays and the $\eta$, $\eta^\prime$ Quark Content,
  Phys.\ Rev.\ D {\bf 41}, 1389 (1990).

\bibitem{Godfrey:2015dia}
  S.~Godfrey and K.~Moats,
  Bottomonium Mesons and Strategies for their Observation,
  Phys.\ Rev.\ D {\bf 92}, 054034 (2015).



\bibitem{Achasov:1994vh}
  N.~N.~Achasov and A.~A.~Kozhevnikov,
Phys.\ Rev.\ D {\bf 49}, 275 (1994).  

\bibitem{Deandrea:2003pv}
  A.~Deandrea, G.~Nardulli and A.~D.~Polosa,
  Phys.\ Rev.\ D {\bf 68}, 034002 (2003)  [hep-ph/0302273].  

\bibitem{Colangelo:2003sa}
  P.~Colangelo, F.~De Fazio and T.~N.~Pham,
  Nonfactorizable contributions in $B$ decays to charmonium: The Case of $B^- \to K^- h_{c}$,
  Phys.\ Rev.\ D {\bf 69}, 054023 (2004).

\bibitem{Lin:1999ad}
  Z.~W.~Lin and C.~M.~Ko,
  A Model for $J/\psi$ absorption in hadronic matter,
  Phys.\ Rev.\ C {\bf 62}, 034903 (2000).

\end{thebibliography}
\end{document}